\documentclass[12pt]{iopart}
\usepackage{iopams}
\usepackage[latin1]{inputenc}
\usepackage{graphicx}
\usepackage{times}
\usepackage{bm}
\usepackage[colorlinks,urlcolor=blue,linkcolor=blue,anchorcolor=blue,citecolor=blue,dvipdfm]{hyperref}
\begin{document}

\title[Global phase diagram, possible chiral SL and topological SC in the triangular Kitaev-Heisenberg model]
{Global phase diagram, possible chiral spin liquid and topological superconductivity in the triangular Kitaev-Heisenberg model}

\author{Kai Li, Shun-Li Yu and Jian-Xin Li}

\address{National Laboratory of Solid State Microstructures
and Department of Physics, Nanjing University, Nanjing 210093,
China\\
Collaborative Innovation Center of Advanced Microstructures,
Nanjing, China

E-mail: dreamkaige@gmail.com, slyu@nju.edu.cn and jxli@nju.edu.cn}

\begin{abstract}
The possible ground states of the undoped and doped Kitaev-Heisenberg model on a triangular lattice are studied. For the undoped system, a combination of the numerical exact diagonalization calculation and the four-sublattice transformation analysis suggests one possible exotic phase and four magnetically ordered phases including a collinear stripe pattern and a noncollinear spiral pattern in the global phase diagram. The exotic phase near the antiferromagnetic (AF) Kitaev point is further investigated by using the Schwinger-fermion mean-field method, and we obtain an energetically favorable $Z_2$ chiral spin liquid with a Chern number $\pm2$ as a promising candidate. At finite doping, we find that the AF Heisenberg coupling supports an $s$-wave or a $d_{x^2-y^2}+id_{xy}$-wave superconductivity (SC), while the AF and the ferromagnetic Kitaev interactions favor a $d_{x^2-y^2}+id_{xy}$-wave SC and a time-reversal invariant topological $p$-wave SC, respectively. Possible experimental realizations and related candidate materials are also discussed.
\end{abstract}


\maketitle
\section{Introduction \label{intro}}
Recently, there has been enormous interest in the physics of the spin-1/2 Kitaev model on a honeycomb lattice \cite{Kitaev2006}, which has an exact $Z_2$ spin-liquid (SL) ground state (GS) supporting fractionalized excitations. One possible route to realize this highly anisotropic spin model is to include a strong relativistic spin-orbit coupling (SOC) in Mott insulators \cite{Khaliullin2005, Khaliullin2009}. Indeed, the interplay of SOC and electron interactions \cite{Balents2010, Rachel2010, Lee2011, Assaad2011, Yu2011} gives rise to many novel phases \cite{Rachel2012, Fiete2012a, Fiete2012b, Fiete2013, Fiete2014}, especially for the so-called relativistic Mott insulators (RMIs) whose physics may drastically differ from that of Mott insulators with weak SOC (e.g., cuprates) \cite{Khaliullin2009, Balents2010, Comin2013, Nagaosa2000}. Of particular interest is the $5d$ transition metal oxides, such as iridates A$_2$IrO$_3$ (A= Na, Li) \cite{Katukuri2014, Gretarsson2013, Singh2012}, where Na$_2$IrO$_3$ is interpreted as a novel RMI \cite{Comin2012} and may also host the quantum spin Hall effect \cite{Shitade2009, Kim2012}. The Kitaev-Heisenberg (KH) model on a honeycomb lattice, which has a rich phase diagram containing unconventional magnetic as well as the Kitaev SL phases \cite{Khaliullin2013, Niu2013}, has been proposed to capture the low-energy properties of A$_2$IrO$_3$ \cite{Khaliullin2010, Kimchi2011}. Meanwhile, experiments confirm a long-range zigzag spin order in Na$_2$IrO$_3$ \cite{Singh2010, Ye2012, Liu2011, Choi2012}, which is a natural GS of the KH model \cite{Khaliullin2013}. In addition, there are also studies on the $4d$ compound Li$_2$RhO$_3$, suggesting Li$_2$RhO$_3$ as a possible RMI with a spin-glass GS \cite{Luo2013}. Theoretical studies also show that carrier doping into RMIs can induce unconventional superconducting pairings as well as the topological superconductivity (SC) \cite{Khaliullin2004, Hyart2012, You2012, Okamoto2013a, Okamoto2013b, Honerkamp2014}.

In fact, the KH model can be generalized to the triangular lattice \cite{Rousochatzakis2012, Vishwanath2014}. Similar to the microscopic origin of the honeycomb KH model for A$_2$IrO$_3$, the triangular KH model can emerge from a class of ABO$_2$ (where A and B are alkali and transition metal ions, respectively) type layered compounds \cite{Khaliullin2005, Khaliullin2009}, due to the joint effect of strong SOC, Coulomb interaction, orbital degeneracy, $t_{2g}^5$ configuration, and $90^\circ$-bonding geometry \cite{Khaliullin2013, Khaliullin2010}. However, to the best of our knowledge, no exact solution of the spin-1/2 Kitaev model on a triangular lattice has been obtained so far. Therefore, it remains conceptually interesting to investigate whether a SL could exist as a possible GS of the triangular KH model.

In this paper, by combining the numerical exact diagonalization (ED) calculation with the four-sublattice transformation (FST) \cite{Khaliullin2005, Khaliullin2013, Khaliullin2010} analysis, we demonstrate one possible exotic phase near the antiferromagnetic (AF) Kitaev point and four magnetically ordered phases including a collinear stripe pattern and a noncollinear spiral pattern in the global phase diagram. For the exotic phase, resorting to the Schwinger-fermion mean-field (MF) method \cite{Affleck1988, Nayak2011}, we find two local minimum solutions with an $s$-wave and a $d+id$-wave pairings, respectively, where the latter has a lower MF energy and is further identified as a $Z_2$ chiral SL state with a Chern number $\pm2$. The effect of finite hole-doping is analyzed by using the slave-boson MF theory, and a time-reversal (TR) invariant topological $p$-wave SC, an $s$-wave SC, and a $d_{x^2-y^2}+id_{xy}$-wave SC are found in the phase diagrams.

\section{Model and exact results}
\begin{figure}[h]
\centering
\includegraphics[width=0.8\textwidth]{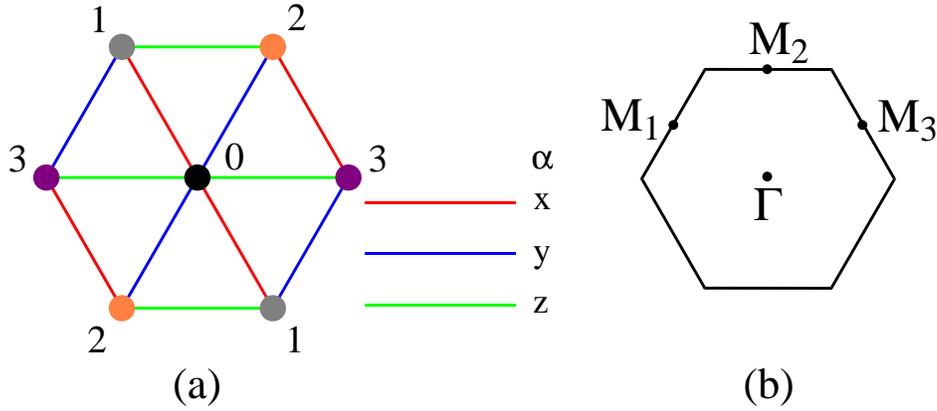} \caption{\label{fig:lattice} (a) Three different directions of the nearest-neighbor bonds on the triangular lattice, namely $\alpha=x,y,z$ colored red, blue, and green, respectively. The numbers 0,1,2,3 label the four sublattices realizing the four-sublattice transformation. (b) The four time-reversal invariant points in the Brillouin zone are $\Gamma,M_1,M_2,$ and $M_3$.}
\end{figure}

Let us begin with the spin-1/2 KH model defined on a triangular lattice
\begin{equation}
H_{KH}=\sum_{\langle i,j\rangle}(-J_{K}S_{i}^{\alpha_{ij}} S_{j}^{\alpha_{ij}}+J_H\bm{S}_{i} \cdot \bm{S}_{j}), \label{eq:KH}
\end{equation}
where the index $\alpha_{ij}$ takes values $x,y,$ or $z$ depending on the direction of the nearest-neighbor (NN) bond $\langle i,j\rangle$ [see figure \ref{fig:lattice}(a)]. This model consists of spin-anisotropic Kitaev interactions (the first term) and spin-isotropic Heisenberg interactions (the second term).

For convenience, we parametrize the coupling constants in equation (\ref{eq:KH}) by introducing the energy scale $A=\sqrt{J_K^2+J_H^2}$ and the angle $\phi$ via $J_K=A\sin\phi$ and $J_H=A\cos\phi$, and let $\phi$ vary from $0$ to $2\pi$ to cover the global phase diagram.

\begin{figure}[h]
\centering
\includegraphics[width=0.8\textwidth]{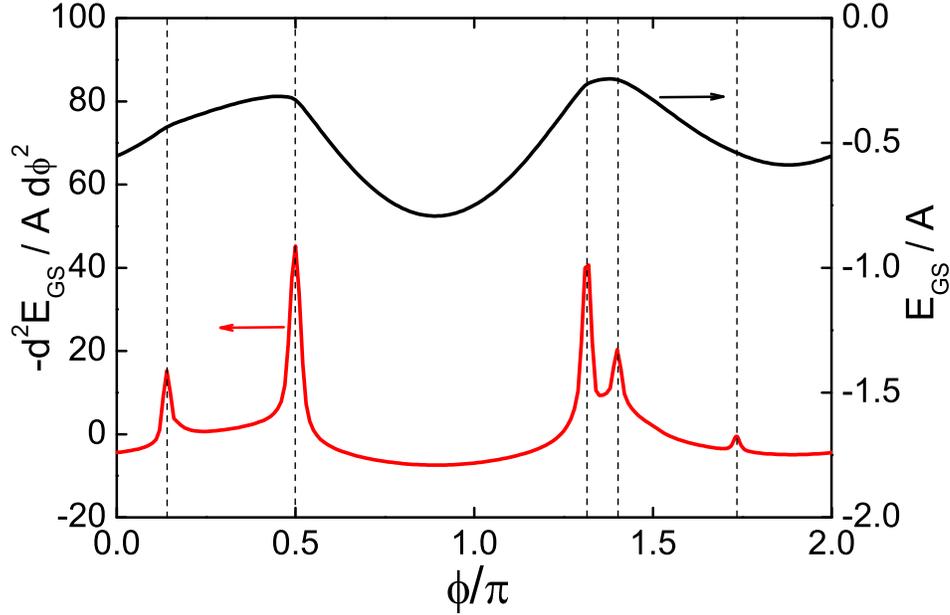} \caption{\label{fig:energy} Lanczos exact diagonalization results for the spin Hamiltonian (\ref{eq:KH}): Ground-state energy $E_\texttt{GS}/A$ per site (black curve) and its second derivative $-d^2E_\texttt{GS}/Ad\phi^2$ (red curve). The dashed lines correspond to the singularities of the ground-state
energy as a function of $\phi$, revealing the phase transitions.}
\end{figure}

\begin{figure}[h]
\centering
\includegraphics[width=0.8\textwidth]{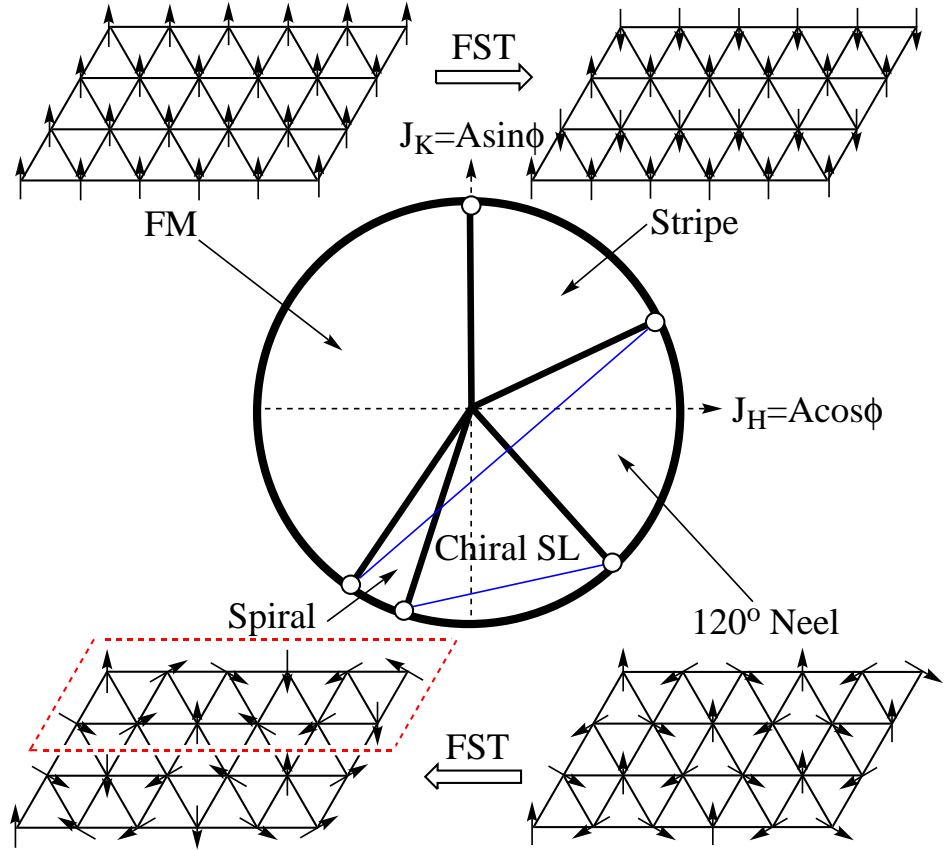} \caption{\label{fig:spin} The global phase diagram of the undoped triangular Kitaev-Heisenberg model containing one SL and four magnetically ordered phases. The transition points (open dots on the $\phi$ circle) are obtained by the exact diagonalization calculation (see figure \ref{fig:energy}). The blue lines inside the circle connect two pairs of transition points that are exactly related by the four-sublattice transformation (see text). The collinear spins in the FM or the stripe pattern (top) lie along the spin $z$ axis, and the noncollinear spins in the $120^\circ$ Neel or the spiral pattern (bottom) are constrained to the spin $(x,y)$ plane. For the spiral phase (bottom-left), the $12$-site magnetic unit cell is enclosed by the red dashed parallelogram.}
\end{figure}

To detect the quantum phase transitions, we perform a Lanczos ED calculation of the GS energy of the Hamiltonian (\ref{eq:KH}) on a 24-site cluster with periodic boundary conditions, and the results are presented in figure \ref{fig:energy}. As indicated by the dashed lines in figure \ref{fig:energy}, the second derivative of the GS energy with respect to $\phi$ reveals five distinct phases separated by five transition points $\phi\approx0.14\pi, 0.5\pi, 1.31\pi, 1.4\pi$, and $1.73\pi$, best visualized using the $\phi$ circle as in figure \ref{fig:spin}.

At $\phi=0$, we are left with the AF Heisenberg model ($J_K=0, J_H>0$) exhibiting the $120^\circ$ Neel order \cite{White2007, Capriotti1999, Bernu1994, Huse1988}; at $\phi=\pi$, equation (\ref{eq:KH}) corresponds to the ferromagnetic (FM) Heisenberg model ($J_K=0, J_H<0$) with a FM GS. In addition to these well-known phases, we observe two more magnetic orders by using the so-called FST approach, which is in fact a spin-rotation transformation. Specifically, it is instructive to divide the triangular lattice into four sublattices [see figure \ref{fig:lattice}(a)] and introduce the rotated spin operators $\widetilde{\bm{S}}$: $\widetilde{\bm{S}}_0=\bm{S}_0$ in the sublattice 0, $\widetilde{\bm{S}}_1=(S_1^x,-S_1^y,-S_1^z)$ in the sublattice 1, $\widetilde{\bm{S}}_2=(-S_2^x,S_2^y,-S_2^z)$ in the sublattice 2, and $\widetilde{\bm{S}}_3=(-S_3^x,-S_3^y,S_3^z)$ in the sublattice 3. This transformation results in the new spin-$\widetilde{\bm{S}}$ Hamiltonian of the same form as equation (\ref{eq:KH}), but with the effective couplings $\widetilde{J}_K=J_K-2J_H$ and $\widetilde{J}_H=-J_H$. For the angles, the mapping reads as $\tan\widetilde{\phi}=-\tan\phi+2$.

Therefore, at $\phi=\arctan2$ and $\arctan2+\pi$, FST maps equation (\ref{eq:KH}) to a FM and an AF Heisenberg Hamiltonians for $\widetilde{\bm{S}}$, respectively. Thus, transforming back to the original spin basis,
we obtain a collinear stripe order at $\phi=\arctan2$ and a noncollinear spiral order at $\phi=\arctan2+\pi$, as shown in figure \ref{fig:spin}, where the spiral pattern has an enlarged magnetic unit cell containing $12$ lattice sites.

Here, we would like to point out that the ED results for both the
two pairs of transition points $(0.14\pi, 1.31\pi)$ and
$(1.4\pi,1.73\pi)$ match the FST mapping
$\tan\widetilde{\phi}=-\tan\phi+2$ very well (as indicated by the
blue lines inside the $\phi$ circle, see figure \ref{fig:spin}),
and the isolated transition point $\phi=0.5\pi$ is also consistent
with FST, i.e., it is mapped to itself under FST.

Thus far, the numerical ED together with the analytical FST has
revealed four magnetically ordered phases in the global phase
diagram. However, we are still left with one possible exotic phase
(i.e., the phase corresponds to the $1.4\pi<\phi<1.73\pi$ arc of
the $\phi$ circle) near the AF Kitaev point (i.e., the point at
$\phi=3\pi/2$).

\section{Possible spin liquid phase in the region $1.4\pi<\phi<1.73\pi$}
Under FST, it can be seen that each point inside the region $1.4\pi<\phi<1.73\pi$ is mapped to another point \emph{still} inside this region, where the AF Kitaev point ($\phi=3\pi/2$) is exactly invariant. Therefore, if the exotic phase ($1.4\pi<\phi<1.73\pi$) is magnetically ordered, it would be somewhat subtle in the sense that the corresponding spin configuration must be invariant under FST. An alternative is a magnetically disordered state that may be favored by the frustration and quantum fluctuations embedded in equation (\ref{eq:Kitaev}).

As a representative of the phase in the region $1.4\pi<\phi<1.73\pi$ and for simplicity, let us start from the AF Kitaev point of the Hamiltonian (\ref{eq:KH}) to explore the nature of the exotic phase, say
\begin{equation}
H_K=-J_K\sum_{\langle i,j\rangle}S_i^{\alpha_{ij}} S_j^{\alpha_{ij}}, \texttt{ } \texttt{with}\texttt{ } J_K<0.  \label{eq:Kitaev}
\end{equation}
The model (\ref{eq:Kitaev}) is TR invariant. It is also invariant under the lattice-translation and -inversion and under the spin-rotation by $\pi$ about the spin $x$, $y$, or $z$ axes.

Unlike the honeycomb Kitaev model \cite{Kitaev2006}, the triangular Kitaev model (\ref{eq:Kitaev}) can not be solved exactly through the Majorana fermionization of spin-$1/2$ operators \cite{Nussinov2009}. Instead, let us take the standard Schwinger-fermion representation of spin-1/2 operators: $S_i^\alpha=\frac{1}{2}\sum_{\sigma,\sigma'=\uparrow,\downarrow}f_{i\sigma}^\dag\tau_{\sigma\sigma'}^\alpha f_{i\sigma'}$, with fermionic spinons $f_{i\sigma}$ and Pauli matrices $\tau^\alpha$. The physical-spin Hilbert space is then recovered by imposing the on-site constraint $\sum_\sigma f_{i\sigma}^\dag f_{i\sigma}=1$, which can be enforced by a site-independent Lagrange multiplier $\lambda$ at the MF level. Now, the spin quadratic terms in equation (\ref{eq:Kitaev}) can be decoupled into several different channels as
\begin{eqnarray}
S_i^xS_j^x=&&-\frac{1}{8}\sum_\sigma t_{\sigma,ij}^{\dag}t_{-\sigma,ij}+
\frac{1}{8}(-\Delta_{0,ij}^\dag\Delta_{0,ij}\nonumber\\&&
-2\Delta_{1,ij}^\dag\Delta_{1,ij}+2\Delta_{2,ij}^\dag\Delta_{2,ij}+\Delta_{3,ij}^\dag\Delta_{3,ij}),
\nonumber\\
S_i^yS_j^y=&&-\frac{1}{8}\sum_\sigma t_{\sigma,ij}^{\dag}t_{-\sigma,ij}+
\frac{1}{8}(-\Delta_{0,ij}^\dag\Delta_{0,ij}\nonumber\\&&
+2\Delta_{1,ij}^\dag\Delta_{1,ij}-2\Delta_{2,ij}^\dag\Delta_{2,ij}+\Delta_{3,ij}^\dag\Delta_{3,ij}),
\nonumber\\
S_i^zS_j^z=&&-\frac{1}{8}\sum_\sigma t_{\sigma,ij}^{\dag}t_{\sigma,ij}+
\frac{1}{8}(-\Delta_{0,ij}^\dag\Delta_{0,ij}\nonumber\\&&
+2\Delta_{1,ij}^\dag\Delta_{1,ij}+2\Delta_{2,ij}^\dag\Delta_{2,ij}-\Delta_{3,ij}^\dag\Delta_{3,ij}),
\label{eq:decoupling}
\end{eqnarray}
where $t_{\sigma,ij}=f_{i\sigma}^\dag f_{j\sigma}$, $\Delta_{0,ij}=\frac{1}{\sqrt{2}}\sum_{\sigma,\sigma'}f_{i\sigma} \mathbf{i}\tau_{\sigma\sigma'}^2f_{j\sigma'}$ (singlet pairing), $\Delta_{l,ij}=\frac{1}{\sqrt{2}}\sum_{\sigma,\sigma'}f_{i\sigma} [\mathbf{i}\tau^l\tau^2]_{\sigma\sigma'}f_{j\sigma'}$ (triplet pairing), with $\mathbf{i}=\sqrt{-1}$, $l=1,2,3$, and the Pauli matrices $\tau^l$.

To proceed, we perform a MF approximation to equation (\ref{eq:decoupling}) with translationally invariant MF Ans\"{a}tze: $t_{\sigma}^{\alpha_{ij}}=\langle t_{\sigma,ij}\rangle$, $\Delta_\mu^{\alpha_{ij}}=\frac{1}{4\sqrt{2}}\langle \Delta_{\mu,ij}\rangle,\mu=0,1,2,3$. By solving these MF parameters self-consistently, we find two local minima of the MF solutions: (1) $\Delta_0^{\alpha}=0.043\equiv\Delta_0,\Delta_l^{\alpha}=t_{\sigma}^{\alpha}=\lambda=0$, leading to an $s$-wave MF Hamiltonian
\begin{equation}
H_s=-\frac{J_K\Delta_0}{2}\sum_{\langle i,j\rangle}[(f_{i\uparrow}^\dag f_{j\downarrow}^\dag-f_{i\downarrow}^\dag f_{j\uparrow}^\dag)+H.c.] \label{eq:sSL}
\end{equation}
with a pseudo Fermi-surface; and (2) $\Delta_0^z=e^{-\mathbf{i}\frac{2\pi}{3}}\Delta_0^x=e^{-\mathbf{i}\frac{4\pi}{3}}\Delta_0^y=0.041, t_{\sigma}^{\alpha}=0.116\equiv t_0,
\Delta_l^\alpha=\lambda=0$, leading to a $d+id$-wave MF Hamiltonian
\begin{equation}
 H_{d+id}=-\frac{J_K}{8}\sum_{\langle i,j\rangle}[-t_0(f_{i\uparrow}^\dag f_{j\uparrow}+f_{i\downarrow}^\dag f_{j\downarrow})
 +4\Delta_0^{\alpha_{ij}}(f_{i\uparrow}^\dag f_{j\downarrow}^\dag-f_{i\downarrow}^\dag f_{j\uparrow}^\dag)+H.c.] \label{eq:dSL}
\end{equation}
with a finite bulk energy gap $0.044|J_K|$.

We further find that $H_{d+id}$ has a lower MF GS energy ($-0.144|J_K|$ per site) than $H_s$ ($-0.107|J_K|$ per site), indicating that the GS of equation (\ref{eq:dSL}) is energetically favorable at the MF level. Therefore, we will focus on the GS properties of $H_{d+id}$ hereafter. The MF Hamiltonian (\ref{eq:dSL}) preserves the lattice-translation and -inversion and the spin-rotation symmetries, and thus the projected physical spin state also preserves these symmetries, which describes a SL phase.

The gauge structure of the MF Hamiltonian $H_{d+id}$ is described by the so-called invariant gauge group (IGG) \cite{Wen2002}. More precisely, to calculate IGG, it is instructive to introduce the notation $\psi_i=(f_{i\uparrow},f_{i\downarrow}^\dag)^T$ and rewrite equation (\ref{eq:dSL}) as
\begin{equation}
H_{d+id}=\sum_{\langle i,j\rangle}(\psi_i^\dag \chi_{ij}\psi_j+H.c.), \label{eq:MF}
\end{equation}
where $\chi_{ij}=-\frac{J_K}{2}\texttt{Re}(\Delta_0^{\alpha_{ij}})\tau^x+\frac{J_K}{2}\texttt{Im}(\Delta_0^{\alpha_{ij}})\tau^y+\frac{J_Kt_0}{8}\tau^z$. The IGG is now defined as the set formed by all the $SU(2)$ gauge transformations that leave $\chi_{ij}$ unchanged, i.e., IGG$=\{G_i|G_i\chi_{ij}G_j^\dag=\chi_{ij},G_i\in SU(2)\}$. For equation (\ref{eq:MF}), we get IGG$=\{G_i=\pm\mathbf{1}\}$ with the $2\times2$ identity matrix $\mathbf{1}$, indicating that the SL state of equation (\ref{eq:dSL}) is in fact a $Z_2$ SL.

The $d+id$ MF Hamiltonian (\ref{eq:dSL}) breaks the TR symmetry by itself (i.e., TR: $H_{d+id}\rightarrow H_{d-id}$). However, from a TR breaking MF Hamiltonian alone, we can not infer the TR symmetry violation \cite{Wen2002, Wen1989}. This is because the Schwinger-fermion representation of the spin-1/2 operators enlarges the spin Hilbert space and introduces an $SU(2)$ gauge redundancy. As a result, the MF GS after projection $\mathcal{P}$ (where $\mathcal{P}$ removes the unphysical states containing empty or doubly occupied sites) may have higher symmetries than the original MF Hamiltonian \cite{Wen2002}.

In practice, the TR symmetry is difficult to verify by writing down the projected GS wavefunction of equation (\ref{eq:dSL}) directly, although this can be done in principle. Following reference \cite{Affleck1988}, we introduce the $SU(2)$ gauge-invariant loop variable to diagnose whether or not the TR symmetry is broken \cite{TRwilson}. The loop variable $W_l$ is defined on a closed, oriented loop $l=j_1\rightarrow j_2 \rightarrow \cdots \rightarrow j_n\rightarrow j_1$ as
\begin{equation}
W_l=\texttt{tr}(\chi_{j_1j_2}\chi_{j_2j_3}\cdots\chi_{j_nj_1}), \label{eq:Wl}
\end{equation}
where $\texttt{tr}(\cdot)$ represents the trace of a matrix. If the $W_l$ configuration is changed under the TR transformation, then the TR symmetry would break down. One can show that under TR
\begin{equation}
W_l\rightarrow \texttt{tr}[(-\tau^y\chi_{j_1j_2}\tau^y)(-\tau^y\chi_{j_2j_3}\tau^y)\cdots(-\tau^y\chi_{j_nj_1}\tau^y)]
=(-1)^nW_l. \label{eq:TRWl}
\end{equation}
Thus, it can be seen that for a bipartite lattice (e.g., the square or honeycomb lattice) where all loops $l$ have even length (i.e., $n$ are even), the $W_l$ configuration is invariant under TR, and hence the TR symmetry should be maintained. On the contrary, for a non-bipartite lattice (e.g., the triangular or kagome lattice) containing odd loops $l$ (i.e., loops with odd length), the corresponding loop variables $W_l$ (if \emph{nonzero}) are changed by a sign under TR and hence the $W_l$ configuration would be changed by TR [see equation (\ref{eq:TRWl})], implying the TR symmetry breaking \cite{TRwilson}.

Here, for the $d+id$ MF Hamiltonian (\ref{eq:dSL}) or (\ref{eq:MF}) on a non-bipartite triangular lattice, we find that the loop variable around each triangular plaquette reads
\begin{equation}
W_\triangle=\pm\mathbf{i}3\sqrt{3}J_K^3t_0|\Delta_0^\alpha|^2/32, \label{eq:triWl}
\end{equation}
which is a \emph{nonzero} number. In fact, these nonzero triangular loop variables reflect the intrinsic frustration of a non-bipartite triangular lattice. Thus, the $Z_2$ SL state described by equation (\ref{eq:dSL}) indeed breaks the TR symmetry. In addition, it also breaks the parity symmetry in two spatial dimensions (i.e., reflection about the axis along the $z$ link). Therefore, our $Z_2$ SL state is a chiral SL state that breaks both the TR symmetry and the parity symmetry \cite{Wen1989}.

The $d+id$ MF Hamiltonian (\ref{eq:dSL}) also has nontrivial band topology as can be seen by rewriting it in momentum space (up to nonessential constant terms): $H_{d+id}=\sum_{\bm{k}}(f_{\bm{k}\uparrow}^\dag, f_{-\bm{k}\downarrow})[\bm{h}(\bm{k})\cdot\bm{\tau}](f_{\bm{k}\uparrow}^\dag, f_{-\bm{k}\downarrow})^\dagger$,
where $\bm{h}(\bm{k})=(\texttt{Re}\Delta_{\bm{k}}, -\texttt{Im}\Delta_{\bm{k}}, \varepsilon_{\bm{k}})$, $\varepsilon_{\bm{k}}=\frac{J_Kt_0}{4}[\cos\bm{k}\cdot\bm{a}_1+\cos\bm{k}\cdot\bm{a}_2
+\cos\bm{k}\cdot(\bm{a}_1+\bm{a}_2)]$, $\Delta_{\bm{k}}=-J_K[\Delta_0^z\cos\bm{k}\cdot\bm{a}_1+\Delta_0^x\cos\bm{k}\cdot\bm{a}_2
+\Delta_0^y\cos\bm{k}\cdot(\bm{a}_1+\bm{a}_2)]$, and the lattice unit vectors $\bm{a}_1=(1,0)$ and $\bm{a}_2=(-1/2,\sqrt{3}/2)$ specify the bond directions $z$ and $x$, respectively. We see that the $2\times2$ Hamiltonian $\bm{h}(\bm{k})\cdot\bm{\tau}$ has two no-crossing bands where each band has a Chern number $\pm2$.

\section{Effect of finite doping\label{doped}}

Away from half filling, we now consider the effect of finite hole-doping by adding the NN hopping terms to the undoped triangular KH model (\ref{eq:KH}), i.e.,
\begin{equation}
H_h=-t\sum_{\langle i,j\rangle,\sigma}c_{i\sigma}^\dagger c_{j\sigma}
-J_K\sum_{\langle i,j\rangle}S_i^{\alpha_{ij}} S_j^{\alpha_{ij}}
+J_H\sum_{\langle i,j\rangle}(\bm{S}_i \cdot \bm{S}_j-\frac{1}{4}\hat{n}_i\hat{n}_j)-\mu\sum_i\hat{n}_i, \label{eq:dopedKH}
\end{equation}
where the chemical potential $\mu$ is adjusted such that $\langle\hat{n}_i\rangle=1-\delta$ with the doping level $\delta$ per site. As in the $t$-$J$ model for high-$T_c$ cuprates, here, we consider the case that the double occupancy is prohibited due to the strong onsite repulsive interactions and adopt the slave boson approach $c_{i\sigma}^\dagger=f_{i\sigma}^\dagger b_i$, with additional bosonic holons $b_i$ that are assumed to be condensed, i.e., $b_i\approx b_i^\dagger\approx\sqrt{\langle b_i^\dagger b_i\rangle}=\sqrt{\delta}$. Thus, the quadratic terms in equation (\ref{eq:dopedKH}) are reduced to $H_T=-\delta t\sum_{\langle i,j\rangle,\sigma}f_{i\sigma}^\dagger f_{j\sigma}-\mu\sum_{i,\sigma}f_{i\sigma}^\dagger f_{i\sigma}+\mu(1-\delta)N$. The spin-exchange terms in equation (\ref{eq:dopedKH}) can now be decoupled into both hopping and pairing channels as in equation (\ref{eq:decoupling}). However, because of the presence of the kinetic term $H_T$, the effect of decoupling spin interactions into hopping channels is not expected to qualitatively change the SC phase diagram at reasonably large doping \cite{Hyart2012}. For simplicity, we thus consider only the pairing channels at finite doping \cite{Hyart2012}, say
$S_i^lS_j^l=-\frac{1}{4}(\Delta_{0,ij}^\dag\Delta_{0,ij}+2\Delta_{l,ij}^\dag\Delta_{l,ij}
-\sum_{m=1}^3\Delta_{m,ij}^\dag\Delta_{m,ij})$,
where the corresponding MF parameters are defined as in the undoped case, $l=1,2,3$ respectively corresponds to the spin component $x,y,z$, and the summation over $l$ gives a Heisenberg term $\bm{S}_i\cdot\bm{S}_j-\frac{1}{4}\hat{n}_i\hat{n}_j=-\Delta_{0,ij}^\dag\Delta_{0,ij}$. This slave boson MF approach has also been widely used to study the doped KH model on the honeycomb lattice very recently \cite{Hyart2012, You2012, Okamoto2013a, Okamoto2013b}, and subsequent unbiased numerical methods \cite{Honerkamp2014} have confirmed those MF results at the qualitative level.

\begin{figure}[h]
\centering
\includegraphics[width=0.8\textwidth]{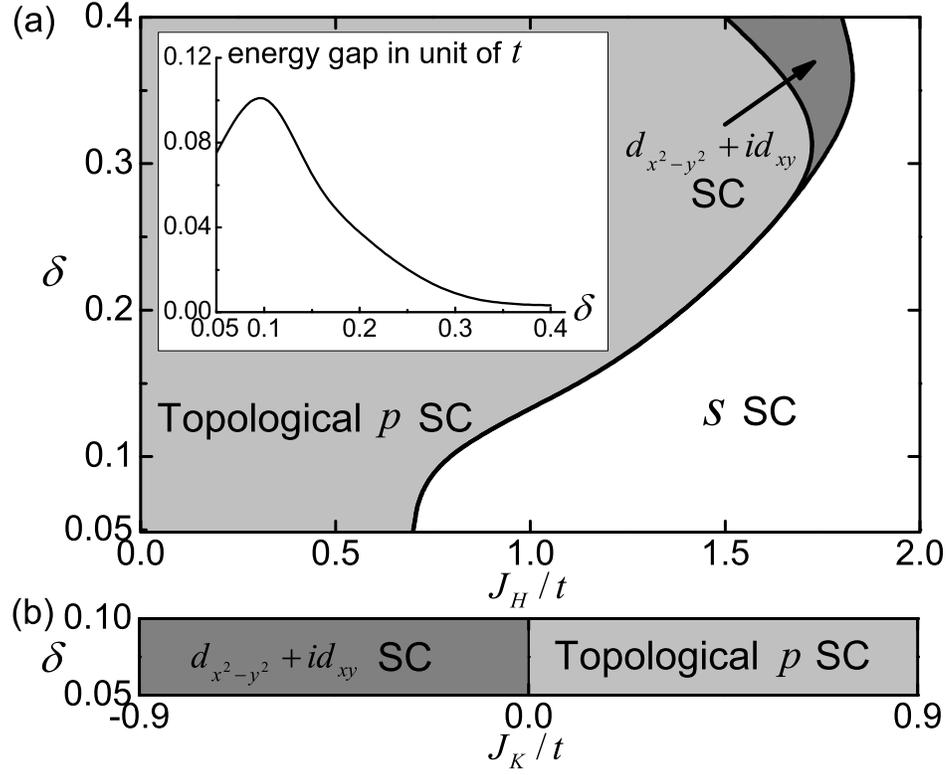} \caption{\label{fig:SC} The mean-field phase diagram of the hole doped triangular Kitaev-Heisenberg model: (a) $J_K=0.5t$ and (b) $J_H=0$. The inset shows the bulk gap of the $p$-wave SC as a function of $\delta$.}
\end{figure}

Here, we focus on the following MF Ans\"{a}tze: (1) $p$-wave pairing, $\Delta_l^x=\Delta_l^y=\Delta_l^z\equiv\Delta_l$; (2) $s$-wave pairing, $\Delta_0^x=\Delta_0^y=\Delta_0^z$; and (3) $d_{x^2-y^2}+id_{xy}$-wave pairing, $e^{-\mathbf{i}\frac{2\pi}{3}}\Delta_0^x=e^{-\mathbf{i}\frac{4\pi}{3}}\Delta_0^y=\Delta_0^z$. Based on these MF Ans\"{a}tze, the resulting phase diagram as a function of $\delta$ and $J_H/J_K$ is shown in figure \ref{fig:SC}, which is qualitatively analogous to the SC phase diagram of the honeycomb KH model at finite doping \cite{Hyart2012, Okamoto2013b, Honerkamp2014}.

We find that (1) the $p$-wave solutions are symmetric $\Delta_1=\Delta_2=\Delta_3$ and pure imaginary, and (2) the $s$-wave solutions are real. Thus, both the $p$-wave SC and the $s$-wave SC are TR invariant. We further note that the $p$-wave SC is fully gapped in the doping interval $0.05\leqslant\delta\leqslant0.4$ [see the inset of figure \ref{fig:SC}(a)], indicating the absence of a topological phase transition in the finite-doping regime. In fact, the topological property of a fully gapped spin-triplet SC with TR symmetry is intimately related to the Fermi-surface topology in the normal state \cite{Sato2009, Qi2011}. More precisely, the $Z_2$ invariant is determined by the parity of the number of TR invariant points \cite{TRpoint} below the Fermi level. Here, the Fermi level is determined by the energy dispersion $\epsilon_{\bm{k}}$ of $H_T$, where $\epsilon_{-\bm{k}}=\epsilon_{\bm{k}}$ due to the lattice-inversion and TR symmetries of $H_T$. For the $p$-wave SC in the doping interval $0.05\leqslant\delta\leqslant0.4$, we find that there is only one TR invariant point $\Gamma$ below the Fermi level. According to the above criterion, the value of the $Z_2$ invariant is odd and the $p$-wave SC is thus topologically nontrivial \cite{Sato2009, Qi2011}.

\begin{figure}[h]
\centering
\includegraphics[width=0.8\textwidth]{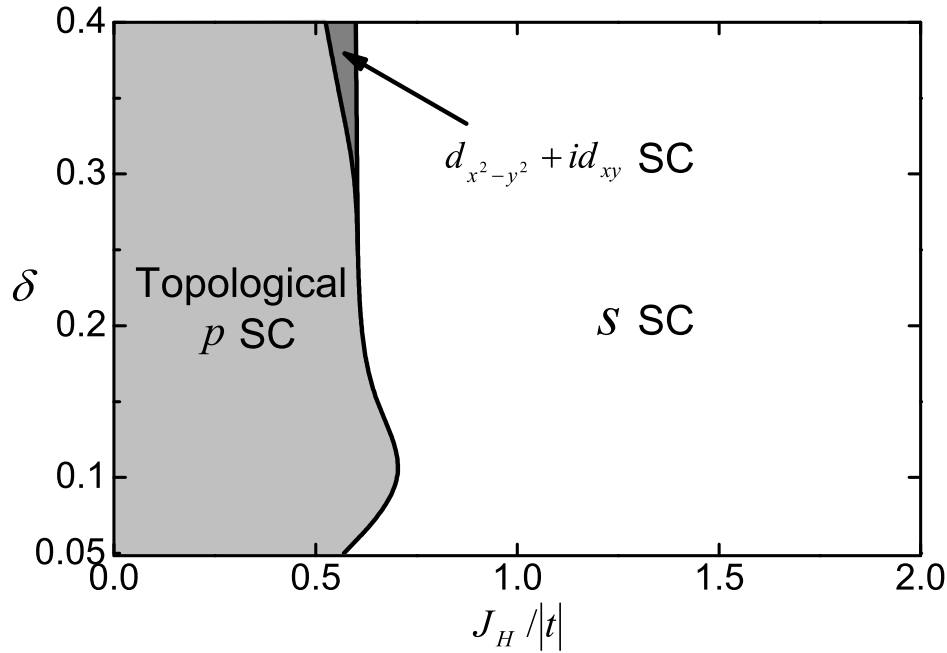} \caption{\label{fig:edoped} The schematic phase diagram of the electron doped triangular Kitaev-Heisenberg model at $J_K=0.5|t|$, which is qualitatively similar to but quantitatively different from the case of hole doping [figure \ref{fig:SC}(a)].}
\end{figure}

The above results focus on the effects of hole doping. In the case
of electron doping, if one wants to study the effects still in the
context of $t$-$J$-type model like equation (\ref{eq:dopedKH}),
technically one needs to perform a particle-hole transformation
$c_{i,\sigma}\rightarrow c_{i,-\sigma}^\dagger$ such that the
constraint now means no double occupancy of holes under electron
doping (or equivalently no empty sites of electrons). The
resulting electron doped Hamiltonian then takes the form
\begin{equation}
H_e=t\sum_{\langle i,j\rangle,\sigma}c_{i\sigma}^\dagger c_{j\sigma}
-J_K\sum_{\langle i,j\rangle}S_i^{\alpha_{ij}} S_j^{\alpha_{ij}}
+J_H\sum_{\langle i,j\rangle}(\bm{S}_i \cdot \bm{S}_j-\frac{1}{4}\hat{n}_i\hat{n}_j)-\mu\sum_i\hat{n}_i. \label{eq:edopedKH}
\end{equation}
Note that as compared with equation (\ref{eq:dopedKH}), the minus
sign before the hopping $t$ disappears and the parameter $t$ is
positive here. The operator $c_{i\sigma}^\dagger$ in equation
(\ref{eq:edopedKH}) now creates a hole instead of an electron, and
the corresponding number operator $\hat{n}_i$ now denotes the
number of holes. And the chemical potential $\mu$ is adjusted such
that $\langle\hat{n}_i\rangle=1-\delta$ with the electron doping
level $\delta>0$ from half filling. Thus, the subsequent
mathematical procedure to solve the electron doped Hamiltonian
(\ref{eq:edopedKH}) is the same as that for the hole doped
Hamiltonian (\ref{eq:dopedKH}) by using the slave boson formalism.
Now the $p$-wave SC, the $s$-wave SC, and the
$d_{x^2-y^2}+id_{xy}$-wave SC MF Ans\"{a}tze under electron doping
are defined as the same form as those under hole doping. Based on
those MF Ans\"{a}tze, a self-consistent calculation yields the SC
phase diagram of electron doping, as shown in figure
\ref{fig:edoped}. We find that both the $p$-wave SC and the
$s$-wave SC are TR invariant, and the $p$-wave SC is fully gapped
in the electron doping interval $0.05\leqslant\delta\leqslant0.4$.
We further note that there are three TR invariant points [say
$M_1,M_2,$ and $M_3$ as shown in figure \ref{fig:lattice}(b)]
below the Fermi level for the $p$-wave SC, indicating that it is
also topologically nontrivial \cite{Sato2009, Qi2011} under
electron doping.

It is noteworthy that the electron doped phase diagram (figure
\ref{fig:edoped}) is qualitatively similar to but quantitatively
different from the hole doped phase diagram [figure
\ref{fig:SC}(a)]. As shown, there also exist three SC phases
which are the same as those under hole doping, but the region of the
triplet SC phase is strongly suppressed and the region of the
singlet SC phase is considerably enlarged. This asymmetry between
the electron doped and the hole doped phase diagrams is due to the
absence of particle-hole symmetry of model (\ref{eq:dopedKH})
on the triangular lattice.

\section{Discussion and conclusion\label{summary}}

Based on the Schwinger-fermion mean-field theory, we show that the
possible candidate phase near the AF Kitaev point for the undoped
triangular Kitaev-Heisenberg model is a gapped $Z_2$ chiral spin
liquid. We note that reference \cite{Rousochatzakis2012} suggests
a nematic order near the AF Kitaev point. Since the investigation
in reference \cite{Rousochatzakis2012} is based on a classical
triangular Kitaev-Heisenberg model, this difference may be due to
the different treatments of the spin dynamics, and whether the
quantum fluctuations would distort the classical nematic order is
waited for further studies.  On the other hand, the $120^\circ$
Neel order has been shown to be destabilized by an infinitesimal
Kitaev term in reference \cite{Rousochatzakis2012} with the Monte
Carlo simulations. The resulting phase consists of non-coplanar
spin patterns remaining locally close to the $120^\circ$ pattern
but distorted at larger distances, which hosts $Z_2$ vortices. For
the local feature of the spin correlations, our result is
consistent with that in reference \cite{Rousochatzakis2012}.
Besides, the magnetically ordered phases such as the collinear
stripe and the noncollinear spiral patterns obtained in our work
are consistent with those in reference \cite{Rousochatzakis2012}.

It is also instructive to compare our results for the undoped triangular Kitaev-Heisenberg model with the honeycomb Kitaev model which has an exact solution. The broken time-reversal symmetry of our $Z_2$ chiral spin liquid stems from the existence of triangular loops $\triangle$ with odd length and the corresponding nonzero loop variables equation (\ref{eq:triWl}). A similar reason also leads to the emergence of an exact chiral spin liquid state in the Kitaev model defined on a decorated honeycomb lattice which also contains odd triangular loops \cite{HYao2007}. By contrast, the exact Kitaev spin liquid in a bipartite honeycomb lattice is time-reversal invariant since all loops there have even length and hence the corresponding loop variables (which are called Wilson loops in \cite{Kitaev2006}) are invariant under the time-reversal transformation \cite{TRwilson}. It is also known that the Kitaev spin liquid in the honeycomb lattice is insensitive to the signs of coupling constants, which is again due to the bipartite nature of the honeycomb lattice: Reversing those signs is simply equivalent to a gauge transformation \cite{Kitaev2006}, which will not affect the physical spin liquid state. This is in sharp contrast to the Kitaev model on a non-bipartite triangular lattice, where we find that the FM Kitaev point (i.e., the point at $\phi=\pi/2$) itself is a critical point separating the stripe and the FM phases while the AF Kitaev point represents the exotic phase which is suggested as a $Z_2$ chiral spin liquid.

In conclusion, we investigate the ground-state phase diagrams of the undoped and doped Kitaev-Heisenberg model on a triangular lattice. In the undoped case, a numerical exact diagonalization calculation shows five transition points separating five distinct phases in the global phase diagram, where two phases are known to be the FM order and the $120^\circ$ Neel order. The four-sublattice transformation sheds further light on the nature of the phase diagram and reveals other two magnetic orders, namely, a collinear stripe pattern and a noncollinear spiral pattern. Lastly, based on the Schwinger-fermion mean-field theory, an energetically favorable $Z_2$ chiral spin liquid with a Chern number $\pm2$ is proposed as a potential candidate for the exotic phase near the AF Kitaev point. Yet, finite hole-doping induces an $s$-wave, a $d_{x^2-y^2}+id_{xy}$-wave, and a time-reversal invariant topological $p$-wave superconducting states.

Finally, a few remarks are in order concerning the possible experimental realization of our results and the relevant materials. The recently synthesized iridate Ba$_3$IrTi$_2$O$_9$ \cite{Dey2012} consisting of a layered triangular arrangement of Ir$^{4+}$ ions with an effective magnetic moment $J_{\texttt{eff}}=\frac{1}{2}$ possesses almost all the necessary ingredients for Kitaev-type exchange couplings (as mentioned in section \ref{intro}), and thus it is a promising candidate for a microscopic realization of the triangular Kitaev-Heisenberg model. Experimental measurements on the Ba$_3$IrTi$_2$O$_9$ show no magnetic ordering down to 0.35K and suggest that Ba$_3$IrTi$_2$O$_9$ probably hosts a spin liquid ground state \cite{Dey2012}, and hence the proposed $Z_2$ chiral spin liquid phase may be observed in this material. In addition, a class of ABO$_2$ (where A and B are alkali and transition metal ions, respectively) type transition metal compounds \cite{Khaliullin2005, Khaliullin2009} and the possible material Na$_x$IrO$_2$ \cite{Vishwanath2014} may be potential candidates for realizing the triangular Kitaev-Heisenberg model. It was also found very recently that both Pt or Pd intercalations and substitutions of layered iridium ditelluride IrTe$_2$ could induce bulk superconductivity with $T_c$ up to $\sim 3$K \cite{yang2012}, and the spin-orbit coupling in these compounds is expected to be strong due to the large atomic numbers of both Ir and Te \cite{yang2012, nlwang}. Thus, a layered IrTe$_2$ with Pt or Pd intercalations and substitutions may be a candidate for the unconventional superconductivity proposed here.

\textit{Note added}. Upon the completion of this manuscript, we
became aware of an independent work \cite{Trebst2014} containing
related results on the undoped case. The four magnetically ordered
phases obtained here are consistent with those in reference \cite{Trebst2014}.
However, reference \cite{Trebst2014} starts with a classical
ground state at the AF Kitaev point and further includes the
effects of quantum fluctuations via a numerical analysis,
consequently it suggests a nematic phase. Therefore, the issue
related to this difference is whether the quantum fluctuations are
strong enough to spoil the classical nematic order into a
disordered phase (e.g., a quantum spin liquid).

\section*{Acknowledgements}

This work was supported by the National Natural Science Foundation of China (91021001, 11190023 and 11204125) and the Ministry of Science and Technology of China (973 Project Grants No.2011CB922101 and No. 2011CB605902).

\renewcommand{\theequation}{A.\arabic{equation}}
\setcounter{equation}{0}

\section*{Appendix. Mean-field Hamiltonians \label{Hk}}
In this Appendix we provide some details of the mean-field calculations in the momentum space. After the mean-field decoupling, in both the undoped and doped cases, the Fourier transformed mean-field Hamiltonians with translationally invariant Ans\"{a}tze take the following form

\begin{eqnarray}
H_{MF}=&&\sum_{\bm{k},\sigma}\varepsilon_{\bm{k}\sigma}f_{\bm{k}\sigma}^\dag f_{\bm{k}\sigma}
+\sum_{\bm{k}}(\Delta_{\bm{k}\uparrow}f_{\bm{k}\uparrow}^\dag f_{-\bm{k}\uparrow}^\dag+\Delta_{\bm{k}\downarrow}f_{\bm{k}\downarrow}^\dag f_{-\bm{k}\downarrow}^\dag\nonumber\\&&
+\Delta_{\bm{k}}^tf_{\bm{k}\uparrow}^\dag f_{-\bm{k}\downarrow}^\dag+\Delta_{\bm{k}}^sf_{\bm{k}\uparrow}^\dag f_{-\bm{k}\downarrow}^\dag+H.c.)+E_0 \label{eq:MFHk}
\end{eqnarray}
with the triplet pairing functions $(\Delta_{\bm{k}\uparrow},\Delta_{\bm{k}\downarrow},\Delta_{\bm{k}}^t)$, the singlet pairing function $\Delta_{\bm{k}}^s$, and a constant term $E_0$. Specifically, in the undoped case, we have
\begin{eqnarray*}
\varepsilon_{\bm{k}\sigma}&=&\frac{J_K}{4}[t^z_\sigma\cos\bm{k}\cdot\bm{a}_1+t^x_\sigma\cos\bm{k}\cdot\bm{a}_2
+t^y_\sigma\cos\bm{k}\cdot(\bm{a}_1+\bm{a}_2)]-\lambda,
\\
\Delta_{\bm{k}\uparrow}&=&-\mathbf{i}J_K[\Delta^z_1\sin\bm{k}\cdot\bm{a}_1-\Delta^x_1\sin\bm{k}\cdot\bm{a}_2
-\Delta^y_1\sin\bm{k}\cdot(\bm{a}_1+\bm{a}_2)]\\&&+J_K[\Delta^z_2\sin\bm{k}\cdot\bm{a}_1+\Delta^x_2\sin\bm{k}\cdot\bm{a}_2
+\Delta^y_2\sin\bm{k}\cdot(\bm{a}_1+\bm{a}_2)],
\\
\Delta_{\bm{k}\downarrow}&=&\mathbf{i}J_K[\Delta^z_1\sin\bm{k}\cdot\bm{a}_1-\Delta^x_1\sin\bm{k}\cdot\bm{a}_2
-\Delta^y_1\sin\bm{k}\cdot(\bm{a}_1+\bm{a}_2)]\\&&+J_K[\Delta^z_2\sin\bm{k}\cdot\bm{a}_1+\Delta^x_2\sin\bm{k}\cdot\bm{a}_2
+\Delta^y_2\sin\bm{k}\cdot(\bm{a}_1+\bm{a}_2)],
\\
\Delta_{\bm{k}}^t&=&-\mathbf{i}J_K[\Delta^z_3\sin\bm{k}\cdot\bm{a}_1-\Delta^x_3\sin\bm{k}\cdot\bm{a}_2
+\Delta^y_3\sin\bm{k}\cdot(\bm{a}_1+\bm{a}_2)],
\\
\Delta_{\bm{k}}^s&=&-J_K[\Delta^z_0\cos\bm{k}\cdot\bm{a}_1+\Delta^x_0\cos\bm{k}\cdot\bm{a}_2
+\Delta^y_0\cos\bm{k}\cdot(\bm{a}_1+\bm{a}_2)],
\end{eqnarray*}
and the constant term
\begin{eqnarray*}
E_0=&&-\frac{NJ_K}{8}(t_\uparrow^{x*}t^x_\downarrow+t_\uparrow^{y*}t^y_\downarrow+H.c.+|t_\uparrow^z|^2+|t_\downarrow^z|^2)
-4NJ_K\sum_\alpha|\Delta_0^\alpha|^2
\\&&
+8NJ_K\sum_\alpha|\Delta_1^\alpha|^2+8NJ_K\sum_\alpha|\Delta_2^\alpha|^2+4NJ_K\sum_\alpha|\Delta_3^\alpha|^2
\\&&
-16NJ_K|\Delta_1^x|^2-16NJ_K|\Delta_2^y|^2-8NJ_K|\Delta_3^z|^2+N\lambda.
\end{eqnarray*}
While in the finite doped case, we have
\begin{eqnarray*}
\varepsilon_{\bm{k}\sigma}&=&-2\delta t[\cos\bm{k}\cdot\bm{a}_1+\cos\bm{k}\cdot\bm{a}_2
+\cos\bm{k}\cdot(\bm{a}_1+\bm{a}_2)]-\mu,
\\
\Delta_{\bm{k}\uparrow}&=&-\mathbf{i}J_K\Delta_1[\sin\bm{k}\cdot\bm{a}_1-\sin\bm{k}\cdot\bm{a}_2
-\sin\bm{k}\cdot(\bm{a}_1+\bm{a}_2)]\\&&+J_K\Delta_2[\sin\bm{k}\cdot\bm{a}_1+\sin\bm{k}\cdot\bm{a}_2
+\sin\bm{k}\cdot(\bm{a}_1+\bm{a}_2)],
\\
\Delta_{\bm{k}\downarrow}&=&\mathbf{i}J_K\Delta_1[\sin\bm{k}\cdot\bm{a}_1-\sin\bm{k}\cdot\bm{a}_2
-\sin\bm{k}\cdot(\bm{a}_1+\bm{a}_2)]\\&&+J_K\Delta_2[\sin\bm{k}\cdot\bm{a}_1+\sin\bm{k}\cdot\bm{a}_2
+\sin\bm{k}\cdot(\bm{a}_1+\bm{a}_2)],
\\
\Delta_{\bm{k}}^t&=&-2\mathbf{i}J_K\Delta_3[\sin\bm{k}\cdot\bm{a}_1-\sin\bm{k}\cdot\bm{a}_2
+\sin\bm{k}\cdot(\bm{a}_1+\bm{a}_2)],
\\
\Delta_{\bm{k}}^s&=&2(4J_H-J_K)[\Delta^z_0\cos\bm{k}\cdot\bm{a}_1+\Delta^x_0\cos\bm{k}\cdot\bm{a}_2
+\Delta^y_0\cos\bm{k}\cdot(\bm{a}_1+\bm{a}_2)],
\end{eqnarray*}
and the constant term
\begin{eqnarray*}
E_0=8N(4J_H-J_K)\sum_\alpha|\Delta_0^\alpha|^2+8NJ_K\sum_l|\Delta_l|^2+N\mu(1-\delta).
\end{eqnarray*}

The quadratic Hamiltonian (\ref{eq:MFHk}) can be diagonalized via the Bogoliubov transformation and the resulting fermion spectrum is given by
\begin{eqnarray*}
E_\pm(\bm{k})=&&\frac{1}{2}|\varepsilon_{\bm{k}\uparrow}-\varepsilon_{\bm{k}\downarrow}|
+\sqrt{\frac{1}{4}(\varepsilon_{\bm{k}\uparrow}+\varepsilon_{\bm{k}\downarrow})^2+\frac{1}{2}(A_{\bm{k}}\pm B_{\bm{k}})}
\end{eqnarray*}
with $A_{\bm{k}}=4\sum_\sigma|\Delta_{\bm{k}\sigma}|^2+|\Delta_{\bm{k}}|^2+|\Delta_{-\bm{k}}|^2$, $B_{\bm{k}}=[A_{\bm{k}}^2-64|\Delta_{\bm{k}\uparrow}|^2|\Delta_{\bm{k}\downarrow}|^2
-4|\Delta_{\bm{k}}|^2|\Delta_{-\bm{k}}|^2
-32\texttt{Re}(\Delta_{\bm{k}\uparrow}^*\Delta_{\bm{k}\downarrow}^*\Delta_{\bm{k}}\Delta_{-\bm{k}})]^\frac{1}{2}$, and $\Delta_{\bm{k}}=\Delta_{\bm{k}}^t+\Delta_{\bm{k}}^s$.

And the ground state energy of equation (\ref{eq:MFHk}) is given by
\begin{equation*}
E_g=\frac{1}{2}\sum_{\bm{k},\sigma}\varepsilon_{\bm{k}\sigma}-\frac{1}{2}\sum_{\bm{k}}[E_+(\bm{k})+E_-(\bm{k})]+E_0.
\end{equation*}


\section*{References}
\begin{thebibliography}{99}
\bibitem{Kitaev2006} Kitaev A 2006  \emph{Ann. Phys.} (N.Y.) \textbf{321} 2
\bibitem{Khaliullin2005} Khaliullin G 2005 \emph{Prog. Theor. Phys. Suppl.} \textbf{160} 155
\bibitem{Khaliullin2009} Jackeli G and Khaliullin G 2009 \emph{Phys. Rev. Lett.} \textbf{102} 017205
\bibitem{Balents2010} Pesin D and Balents L 2010 \emph{Nature Physics} \textbf{6} 376
\bibitem{Rachel2010} Rachel S and Hur K Le 2010 \emph{Phys. Rev. B} \textbf{82} 075106
\bibitem{Lee2011} Lee D H 2011 \emph{Phys. Rev. Lett.} \textbf{107} 166806
\bibitem{Assaad2011} Hohenadler M, Lang T C and Assaad F F 2011 \emph{Phys. Rev. Lett.} \textbf{106} 100403
\bibitem{Yu2011} Yu S L, Xie X C and Li J X 2011 \emph{Phys. Rev. Lett.} \textbf{107} 010401
\bibitem{Rachel2012} Reuther J, Thomale R and Rachel S 2012 \emph{Phys. Rev. B} \textbf{86} 155127
\bibitem{Fiete2012a} Kargarian M, Langari A and Fiete G A 2012 \emph{Phys. Rev. B} \textbf{86} 205124
\bibitem{Fiete2012b} R\"{u}egg A and Fiete G A 2012 \emph{Phys. Rev. Lett.} \textbf{108} 046401
\bibitem{Fiete2013} Kargarian M and Fiete G A 2013 \emph{Phys. Rev. Lett.} \textbf{110} 156403
\bibitem{Fiete2014} Maciejko J, Chua V and Fiete G A 2014 \emph{Phys. Rev. Lett.} \textbf{112} 016404
\bibitem{Comin2013} Comin R and Damascelli A 2013 arXiv:1303.1438
\bibitem{Nagaosa2000} Tokura Y and Nagaosa N 2000 \emph{Science} \textbf{288} 462
\bibitem{Katukuri2014} Katukuri V M, Nishimoto S, Yushankhai V, Stoyanova A, Kandpal H, Choi S, Coldea R, Rousochatzakis I, Hozoi L and van den Brink J 2014 \emph{New J. Phys.} \textbf{16} 013056
\bibitem{Gretarsson2013} Gretarsson H, Clancy J P, Liu X, Hill J P, Bozin E, Singh Y, Manni S, Gegenwart P, Kim J, Said A H, Casa D, Gog T, Upton M H, Kim H S, Yu J, Katukuri V M, Hozoi L, van den Brink J and Kim Y J 2013 \emph{Phys. Rev. Lett.} \textbf{110} 076402
\bibitem{Singh2012} Singh Y, Manni S, Reuther J, Berlijn T, Thomale R, Ku W, Trebst S and Gegenwart P 2012 \emph{Phys. Rev. Lett.} \textbf{108} 127203
\bibitem{Comin2012} Comin R, Levy G, Ludbrook B, Zhu Z-H, Veenstra C N, Rosen J A, Singh Y, Gegenwart P, Stricker D, Hancock J N, Marel D van der, Elfimov I S and Damascelli A 2012 \emph{Phys. Rev. Lett.} \textbf{109} 266406
\bibitem{Shitade2009} Shitade A, Katsura H, Kune\v{s} J, Qi X-L, Zhang S-C and Nagaosa N 2009 \emph{Phys. Rev. Lett.} \textbf{102} 256403
\bibitem{Kim2012} Kim C H, Kim H S, Jeong H, Jin H and Yu J 2012 \emph{Phys. Rev. Lett.} \textbf{108} 106401
\bibitem{Khaliullin2013} Chaloupka J, Jackeli G and Khaliullin G 2013 \emph{Phys. Rev. Lett.} \textbf{110} 097204
\bibitem{Niu2013} Yu Y, Liang L, Niu Q and Qin S 2013 \emph{Phys. Rev. B} \textbf{87} 041107(R)
\bibitem{Khaliullin2010} Chaloupka J, Jackeli G and Khaliullin G 2010 \emph{Phys. Rev. Lett.} \textbf{105} 027204
\bibitem{Kimchi2011} Kimchi I and You Y Z 2011 \emph{Phys. Rev. B} \textbf{84} 180407(R)
\bibitem{Singh2010} Singh Y and Gegenwart P 2010 \emph{Phys. Rev. B} \textbf{82} 064412
\bibitem{Ye2012} Ye F, Chi S, Cao H, Chakoumakos B C, Fernandez-Baca J A, Custelcean R, Qi T F, Korneta O B and Cao G 2012 \emph{Phys. Rev. B} \textbf{85} 180403(R)
\bibitem{Liu2011} Liu X, Berlijn T, Yin W-G, Ku W, Tsvelik A, Kim Y-J, Gretarsson H, Singh Y, Gegenwart P and Hill J P 2011 \emph{Phys. Rev. B} \textbf{83} 220403(R)
\bibitem{Choi2012} Choi S K, Coldea R, Kolmogorov A N, Lancaster T, Mazin I I, Blundell S J, Radaelli P G, Singh Y, Gegenwart P, Choi K R, Cheong S-W, Baker P J, Stock C and Taylor J 2012 \emph{Phys. Rev. Lett.} \textbf{108} 127204
\bibitem{Luo2013} Luo Y, Cao C, Si B, Li Y, Bao J, Guo H, Yang X, Shen C, Feng C, Dai J, Cao G and Xu Z-A 2013 \emph{Phys. Rev. B} \textbf{87} 161121(R)
\bibitem{Khaliullin2004} Khaliullin G, Koshibae W and Maekawa S 2004 \emph{Phys. Rev. Lett.} \textbf{93} 176401
\bibitem{Hyart2012} Hyart T, Wright A R, Khaliullin G and Rosenow B 2012 \emph{Phys. Rev. B} \textbf{85} 140510(R)
\bibitem{You2012} You Y-Z, Kimchi I and Vishwanath A 2012 \emph{Phys. Rev. B} \textbf{86} 085145
\bibitem{Okamoto2013a} Okamoto S 2013 \emph{Phys. Rev. Lett.} \textbf{110} 066403
\bibitem{Okamoto2013b} Okamoto S 2013 \emph{Phys. Rev. B} \textbf{87} 064508
\bibitem{Honerkamp2014} Scherer D D, Scherer M M, Khaliullin G, Honerkamp C and Rosenow B 2014 \emph{Phys. Rev. B} \textbf{90} 045135
\bibitem{Rousochatzakis2012} Rousochatzakis I, R\"{o}ssler U K, van den Brink J and Daghofer M 2012 arXiv:1209.5895
\bibitem{Vishwanath2014} Kimchi I and Vishwanath A 2014 \emph{Phys. Rev. B} \textbf{89} 014414
\bibitem{Affleck1988} Affleck I, Zou Z, Hsu T and Anderson P W 1988 \emph{Phys. Rev. B} \textbf{38} 745
\bibitem{Nayak2011} Burnell F J and Nayak C 2011 \emph{Phys. Rev. B} \textbf{84} 125125
\bibitem{White2007} White S R and Chernyshev A L 2007 \emph{Phys. Rev. Lett.} \textbf{99} 127004
\bibitem{Capriotti1999} Capriotti L, Trumper A E and Sorella S 1999 \emph{Phys. Rev. Lett.} \textbf{82} 3899
\bibitem{Bernu1994} Bernu B, Lecheminant P, Lhuillier C and Pierre L 1994 \emph{Phys. Rev. B} \textbf{50} 10048
\bibitem{Huse1988} Huse D A and Elser V 1988 \emph{Phys. Rev. Lett.} \textbf{60} 2531
\bibitem{Nussinov2009}For the general construction of exactly solvable Kitaev-type models on various lattices, see, e.g., Nussinov Z and Ortiz G 2009 \emph{Phys. Rev. B} \textbf{79} 214440
\bibitem{Wen2002} Wen X G 2002 \emph{Phys. Rev. B} \textbf{65} 165113
\bibitem{Wen1989} Wen X G, Wilczek F and Zee A 1989 \emph{Phys. Rev. B} \textbf{39} 11413
\bibitem{TRwilson} See \cite{Kitaev2006} for comments on the intimate relations between (broken) TR symmetry and loop variables (which are called Wilson loops in \cite{Kitaev2006}).
\bibitem{Sato2009} Sato M 2009 \emph{Phys. Rev. B} \textbf{79} 214526
\bibitem{Qi2011} Qi X-L and Zhang S-C 2011 \emph{Rev. Mod. Phys.} \textbf{83} 1057
\bibitem{TRpoint}In two dimensions, there are totally four TR invariant points in Brillouin zone, i.e., $(n_1\bm{b}_1+n_2\bm{b}_2)/2$, as labeled by $\Gamma,M_1,M_2,$ and $M_3$ in figure \ref{fig:lattice}(b), where $n_1,n_2=0,1$, and $\bm{b}_1,\bm{b}_2$ are the primitive reciprocal lattice vectors.
\bibitem{HYao2007} Yao H and Kivelson S A 2007 \emph{Phys. Rev. Lett.} \textbf{99} 247203
\bibitem{Dey2012} Dey T, Mahajan A V, Khuntia P, Baenitz M, Koteswararao B and Chou F C 2012 \emph{Phys. Rev. B} \textbf{86} 140405(R)
\bibitem{yang2012} Yang J J, Choi Y J, Oh Y S, Hogan A, Horibe Y, Kim K, Min B I and Cheong S-W 2012 \emph{Phys. Rev. Lett.} \textbf{108} 116402
\bibitem{nlwang} Fang A F, Xu G, Dong T, Zheng P and Wang N L 2013 \emph{Sci. Rep.} \textbf{3} 1153
\bibitem{Trebst2014} Becker M, Hermanns M, Bauer B, Garst M and Trebst S 2014 arXiv:1409.6972

\end{thebibliography}
\end{document}